\documentstyle[12pt,epsfig]{article}
\renewcommand{\theequation}{\arabic{equation}}
\renewcommand{\thesection}{\arabic{section}}
\renewcommand{\thefootnote}{\fnsymbol{footnote}}

\newcommand{\bea}{\begin{eqnarray}}
\newcommand{\ena}{\end{eqnarray}}
\newcommand{\vs}[1]{\vspace{#1 mm}}

\renewcommand{\a}{\alpha}
\renewcommand{\b}{\beta}

\newcommand{\z}{\omega}

\newcommand{\PL}[1]{Phys.\ Lett.\ {\bf #1}}

\newcommand{\PR}[1]{Phys.\ Rev.\ {\bf #1}}

\newcommand{\EPJ}[1]{Eur.\ Phys.\ J.\ {\bf #1}}
\newcommand{\NP}[1]{Nucl.\ Phys.\ {\bf #1}}

\begin{document}
\noindent
\topmargin 0pt
\oddsidemargin 5mm

\begin{titlepage}
\setcounter{page}{0}
\begin{flushright}
May, 2000\\
OU-HET 347\\
hep-ph/0005267\\
\end{flushright}
\vs{4}
\begin{center}
{\Large{\bf  Renormalization group effect and 
a democratic-type neutrino mass matrix}}\\
\vs{4}
{\large  
Takahiro Miura\footnote{e-mail address: 
miura@het.phys.sci.osaka-u.ac.jp}, 
Tetsuo Shindou\footnote{e-mail address: 
shindou@het.phys.sci.osaka-u.ac.jp},
Eiichi Takasugi\footnote{e-mail address: 
takasugi@het.phys.sci.osaka-u.ac.jp} \\
\vs{2}
{\em Department of Physics,
Osaka University \\ Toyonaka, Osaka 560-0043, Japan} \\
\vs{2}
Masaki Yoshimura\footnote{e-mail address:
myv20012@se.ritsumei.ac.jp}\\
\vs{2}
{\em Department of Physics,
Ritsumeikan University \\ Kusatsu, Shiga 525-8577, Japan} }
\end{center}
\vs{6}
\centerline{{\bf Abstract}}  
In our previous paper, we proposed the democratic-type 
neutrino mass matrix which gives interesting 
predictions, $\theta_{23}=-\frac{\pi}4$, 
$|\tan \theta_{12}|=\sqrt{2-3\sin^2 \theta_{13}}$ and 
$\delta=\frac {\pi}2$, where $\theta_{ij}$ is the mixing 
angle between mass eigenstates $\nu_i$ and $\nu_j$, and 
$\delta$ is the CP violation angle in the standard 
parameterization of mixing matrix. In this paper, we 
examined how predictions behave at $m_Z$  
by assuming that they are given at the right-handed 
neutrino mass scale, $M_R$. 
 
\end{titlepage}

\newpage
\renewcommand{\thefootnote}{\arabic{footnote}}
\setcounter{footnote}{0}

\section{Introduction}

In our recent papers[1],[2], we proposed the democratic-type 
mass matrix which contains six real parameters and found that 
this mass matrix predicts  
\bea
\theta_{23}=-\frac{\pi}4\;,\;\delta=\frac {\pi}2\;,
\ena
where $\theta_{23}$ and $\delta$ are the mixing angle between 
the mass eigenstates, $\nu_2$ and $\nu_3$, and the CP violation 
phase, in the parameterization of the mixing matrix 
given in the particle data group[3] (see the matrix given 
in the Appendix A). 

If we take the CHOOZ bound[4], $|V_{13}|<0.16$ or  
$|\sin \theta_{13}|<0.16$, we find almost maximum 
atmospheric neutrino mixing,
\bea
\sin^2 2\theta_{atm}=4|V_{23}|^2(1-|V_{23}|^2)
=1-\sin^4\theta_{13}>0.999\;,
\ena 
where $V$ is the neutrino mixing matrix.
If the experimental data 
turns out to show that $\sin^2 2\theta_{atm}$ is really 
close to unity, our model will become a good candidate. 
Another special feature of the model is the prediction 
of the value of the CP violation phase. Both Dirac CP  
phase ($\delta$) and Majorana CP phases[5] are predicted[1]. 
In particular, the maximal value of the CP violation phase 
$\delta$ is predicted. Our prediction gives the great encouragement 
for experiments to measure the CP violation in the 
oscillation processes[6] in the near future. The 
theoretical study has become an urgent topic. 

In Ref.2, we made a further investigation on the 
democratic-type mass matrix. We constructed $Z_3$ 
invariant Lagrangian with two or three up-type Higgs 
doublets and derived the democratic-type mass matrix 
by using the see-saw mechanism. We also considered 
one up-type Higgs model. By considering the $Z_3$ symmetric 
Lagrangian together with the $Z_2$ invariant $Z_3$ breaking terms, 
we found the further prediction, 
\bea
|\tan \theta_{12}|=\sqrt{2-3\sin^2 \theta_{13}}\;,
\ena
which we shall explain in the next section. 
By using the CHOOZ bound, this relation leads to  
\bea
0.87<\sin^2 2\theta_{sol}=4|V_{11}|^2|V_{12}|^2<\frac89\;.
\ena

In Refs.1 and 2, we assumed that the above predictions 
are valid at the weak scale $m_Z$, although the neutrino 
mass matrix is assumed to be defined at the 
right-handed neutrino mass scale $M_R$. 
The stability of mixing angles 
under the change of energy scale has been discussed[7-10]. 
According to their result, in many occasions,  
the predictions at $m_Z$ are essentially the same as 
those at $M_R$. In some special cases where $m_1\simeq m_2$, 
the prediction of $\sin^2 2\theta_{sol}$ becomes 
unstable. That is, the predicted large value of 
$\sin^2 2\theta_{sol}$ at $M_R$ becomes the small value at 
$m_Z$.  

The purpose of this paper is to examine the stability of 
our predictions. In particular, we are interested in the 
possibility that the large solar neutrino mixing at $M_R$ 
becomes small to be consistent with the small angle MSW 
solution at $m_Z$. We found that the angle can become small, 
but unfortunately this possibility does not realize the 
small angle MSW solution. 

In Sec.2, we briefly explain our model. In Sec.3, we 
analytically examine the renormalization 
effect on the neutrino mass matrix and the effect to 
our predictions.   
The numerical analysis to supplement the discussions in 
Sec.3 is given in Sec.4. In Sec.5, 
the summary is given.

\section{The model} 

We consider the following dimension five Lagrangian 
in the mass eigenstate basis of charged leptons[2]
\bea
{\cal L}_{Y}
=-(m^0_1+\tilde{m}_1)\overline{(\Psi_1)^C} \Psi_1 
                        \frac{H_u H_u}{u^2_u}
  -2\tilde{m}_1\overline{(\Psi_2)^C} \Psi_3 \frac{H_u H_u}{u^2_u}\;,
\ena
where $\tilde m_1$ and $m_1^0$ are real parameters, and 
$u_u$ is the vacuum expectation value of the neutral 
component of the doublet Higgs $H_u$.  
This Lagrangian is invariant under the $Z_3$ transformation 
\bea
\Psi_1 \to \z \Psi_1\;,\;\Psi_2\to \z^2 \Psi_2\;,\;
\Psi_3 \to \Psi_3\;,\;H_u \rightarrow \z^2 H_u\;,
\ena
where, the irreducible representation $\Psi_i$ $(i=1,2,3)$ 
are defined by 
\bea
\Psi_1&=&\frac1{\sqrt 3}(\ell_e+\z^2 \ell_\mu+\z \ell_\tau)\;,
\nonumber\\
\Psi_2&=&\frac1{\sqrt 3}(\ell_e+\z \ell_\mu+\z^2 \ell_\tau)  
\;,\nonumber\\
\Psi_3&=&\frac1{\sqrt 3}(\ell_e+\ell_\mu+\ell_\tau) 
\;.
\ena
The $Z_3$ transformation for $\Psi_i$ is induced by the cyclic 
permutation among $\ell_i$, which  are the 
left-handed lepton doublets defined by $\ell_e=(\nu_{eL}, e_L)^T$ 
and so on. 

Then, we introduce the $Z_3$ symmetry breaking term, but it 
preserves the  $Z_2$ symmetry
\bea
\Psi_1 \to -\Psi_1\;, 
\ena
and all other fields are unchanged. Now, we find
\bea
 {\cal L}_{SB}=   - m^0_2 \overline{(\Psi_2)^C} \Psi_2 
                        \frac{H_u H_u}{u^2_u}  
   -m^0_3 \overline{(\Psi_3)^C} \Psi_3 
                        \frac{H_u H_u}{u_u^2}\;.
\ena
After $H_u$ acquires the vacuum expectation value, the neutrino 
mass term is derived. In the flavor basis, $(\nu_e,\nu_\mu,\nu_\tau)$, 
the mass matrix is given in a democratic-type form[2], 
\bea
m_\nu (M_R)&=&
\frac{m^0_1}3\pmatrix{1&\z^2&\z\cr \z^2&\z&1\cr 
         \z&1&\z^2}+
 \tilde{m}_1\pmatrix{1&0&0\cr 0&\z&0\cr 0&0&\z^2\cr}
\nonumber\\
&&\hskip 2cm+\frac{m^0_2}3 \pmatrix{1&\z&\z^2\cr \z&\z^2&1\cr
\z^2&1&\z\cr}
+\frac{m^0_3}3 \pmatrix{1&1&1\cr 1&1&1\cr 1&1&1\cr}\;,
\ena
where $\z$ is the element of $Z_3$ symmetry and we take 
$\z=\exp(i2\pi/3)$, i.e., $\z^3=1$. 
We consider that this mass matrix is given at the right-handed 
mass scale $M_R$. 

The unitary matrix $V_2$ which diagonalizes $m_\nu (M_R)$ is derived 
in the Appendix A and the result is 
\bea
V_2= \pmatrix{1&0&0\cr0&\z&0\cr 0&0&\z^2}
   \pmatrix{\frac1{\sqrt{3}}&-\sqrt{\frac23}c'&i\sqrt{\frac23}s'\cr 
            \frac 1{\sqrt 3}&\frac1{\sqrt 6}(c'+i\sqrt 3 s')
            &-\frac1{\sqrt 6}(\sqrt 3 c'+is')\cr
            \frac1{\sqrt 3}&\frac1{\sqrt 6}(c'-i\sqrt 3 s')
            &\frac1{\sqrt 6}(\sqrt 3 c'-is')\cr}
   \pmatrix{1&0&0\cr 0&-1&0\cr 0&0&i\cr}\;,
\ena
where $c'=\cos \theta'$ and $s'=\sin \theta'$ and 
\bea
\tan \theta'=\frac{\Delta_-}
{\tilde m_1+\sqrt{\tilde m_1^2+ \Delta_-^2}}\;,
\ena
with $\Delta_-=(m_3^0-m_2^0)/2$.

It should be noted that predictions in Eqs.(1) and (3) 
are derived from $V_2$. The phase in ${\rm diag}(1,-1,i)$
represents the Majorana phase, while the phases in 
${\rm diag}(1,\z,\z^2)$ are the irrelevant phases
which are absorbed by the charged lepton fields.

From our later analysis, there is essentially no effect 
to $V_{13}$. As a result, we can impose the CHOOZ bound, 
$|V_{13}|<0.16$ at $m_Z$. We find 
\bea
|s'|<0.2\;.
\ena 

We define the mass eigenstate neutrinos at $M_R$ as 
$(\nu_1^R,\nu_2^R,\nu_3^R)$ and their masses are 
\bea
m_1^R&=&m_1^0+\tilde m_1\;,\nonumber\\
m_2^R&=&m_2^0+\Delta_- +\sqrt{\tilde m_1^2+ \Delta_-^2}\;,
\nonumber\\
m_3^R&=&m_2^0+\Delta_- -\sqrt{\tilde m_1^2+ \Delta_-^2}\;. 
\ena

We take the convention, $\tilde m_1>0$. Since $m_2^0$ and 
$m_3^0$ are parameters for the symmetry breaking terms, 
we expect that  $\tilde m_1>> |m_2^0|$, $|m_3^0|$.  
Then, we find $m_2^R>0$ and $m_3^R<0$. 
The parameter $\tilde m_1$ controls the overall normalization  
neutrino masses, and $m_1^0$ and $m_2^0$ (or $m_3^0$) 
control the mass of $m_1^R$ and the mass splitting between 
$m_2^R$ and $m_3^R$, while the parameter $\Delta_-=(m_3^0-m_2^0)/2$ 
does the size of $V_{13}$.

\section{The renormalization group analysis}

We consider the renormalization group effect on the 
dimension five interaction in Eqs.(5) and (9) in the MSSM model. 
The general feature of the stability of mixing angles 
has been extensively discussed[7-10]. Here, 
we take the special mass matrix, 
the democratic-type mass matrix and examine 
the stability in detail. 

\vskip 2mm
\noindent
(3-1) Neutrino mass matrix at $m_Z$

In the basis where charged lepton mass matrix is diagonal and 
thus the Yukawa coupling matrix which induces masses of 
charged leptons is diagonal, 
the neutrino mass matrices at 
$M_R$ and $m_Z$ are related as[7]   
\bea
m_{\nu}(M_R) \simeq  A
\pmatrix{
    1&0&0\cr
   0&\sqrt{\frac{I_{\mu}}{I_{e}}} &0\cr
   0&0&\sqrt{\frac{I_{\tau}}{I_{e}}}\cr}
   m_{\nu}(m_Z)
\pmatrix{
    1&0&0\cr
   0&\sqrt{\frac{I_{\mu}}{I_{e}}} &0\cr
   0&0&\sqrt{\frac{I_{\tau}}{I_{e}}}\cr}\;,
\ena
where 
\bea
I_i = {\rm exp}\Biggl(\frac{1}{8\pi^2}
\int^{{\rm ln} (M_{R})}_{{\rm ln}(m_{z})}y_i^2dt
\Biggr)\;\;\;\;\; (i=e,\mu,\tau ),
\ena
with the Yukawa coupling for charged leptons $y_i$ 
and $A=(I_{e}/I_{\tau})(m_{\nu 33}(M_{R})/m_{\nu 33}(m_Z))$. 

After absorbing $A$ into the overall normalization of  
parameters in $m_\nu(M_R)$ and by using the approximation, 
\bea
\sqrt{\frac{I_{\mu}}{I_{\tau}}}\simeq \sqrt{\frac{I_{e}}
{I_{\tau}}}\simeq \frac{1}{\sqrt{I_{\tau}}}\;,
\ena
we obtain
\bea
m_{\nu}(m_{Z}) =  
\pmatrix{
   1&0&0\cr
   0&1 &0\cr
   0&0&\a \cr}
m_{\nu}(M_{ R}) 
\pmatrix{
   1&0&0\cr
   0&1 &0\cr
   0&0&\a \cr}\;,
\ena
where
\bea
\a \equiv 1/\sqrt{I_{\tau}}
= \Biggl( \frac{m_Z}{M_R} \Biggr)
^{\frac{1}{8\pi^2}(1+\tan^2\b)(m_{\tau}/v)^2} <1\;.
\ena
Here we neglect the radiative correction on $y_{\tau}$, 
and $m_{\tau}$ is the $\tau$ lepton mass, 
$v^2=u^2_u +u^2_d$ and $\tan\b = u_u/u_d$ with
$u_i$ being the vacuum expectation value of 
MSSM Higgs doublet $<H_i>(i=u,d)$. 

Now we define the small parameter $\epsilon =1-\alpha$. 
In order to estimate the value of $\epsilon$, we consider 
the right-handed mass scale $M_R$ and 
the region of $\tan \beta$ as
\bea
M_{R}=10^{13}({\rm GeV})\;,\qquad 2< \tan \beta <60\;.
\ena 
Then, with  $m_Z=91.187({\rm GeV})$, $m_{\tau}=1.777
({\rm GeV})$ and $v=245.4({\rm GeV})$, we find 
\bea
8\times 10^{-5}< \epsilon < 6\times 10^{-2}\;.
\ena

\vskip 2mm
\noindent
(3-2) The diagonalization

By transforming $m_\nu(m_Z)$ in Eq.(18) by $V_2$, we find 
\bea
\tilde m_\nu &\equiv& V_2^T
\pmatrix{1&0&0\cr 0&1&0\cr 0&0&\a\cr}V_2^*
\left (V_2^T m_\nu(M_R)V_2\right )V_2^\dagger
\pmatrix{1&0&0\cr 0&1&0\cr 0&0&\a\cr}V_2
\nonumber\\
&=&(1-\epsilon S^T)D_\nu^R (1-\epsilon S)\;, 
\ena
where 
$D_\nu^R=V_2^T m_\nu(M_R)V_2={\rm diag}(m_1^R,m_2^R,m_3^R)$ 
and 
\bea
S=V_2^\dagger\pmatrix{0&0&0\cr0&0&0\cr0&0&1\cr}V_2 
=\frac13\pmatrix{1&-\frac{1}{\sqrt{2}}ae^{-i\phi_1}&
   i\sqrt{\frac{3}{2}}be^{-i\phi_2}  \cr
  -\frac{1}{\sqrt{2}}a e^{i\phi_1}& \frac{1}{2}a^2&
   -i\frac{\sqrt{3}}{2}ab e^{i(\phi_1-\phi_2)}\cr
  - i\sqrt{\frac{3}{2}}b e^{i\phi_2} &
   i\frac{\sqrt{3}}{2}ab e^{-i(\phi_1-\phi_2)}& 
  \frac{3}{2}b^2 \cr}\;,
\nonumber\\
\ena
where $s'$ and $c'$ are given in Eq.(12), and 
$a$, $b$ and phases $\phi_i$ are defined by 
\bea
a=\sqrt{1+2s'^2}\;,&&\;\; b=\sqrt{1-\frac23s'^2}\;,
\nonumber\\
\tan \phi_1=\sqrt{3}\tan \theta'\;,&&\;
\tan \phi_2=\frac1{\sqrt{3}}\tan \theta'\;.
\ena

By keeping $\epsilon$  
up to the first order, we find
\bea
\tilde m_\nu \simeq  
\pmatrix{(1 -\frac23\epsilon)m_1^R&
 \frac1{3\sqrt{2}} \epsilon a(m_1^Re^{-i\phi_1}
 +m_2^Re^{i\phi_1})
 &-i\epsilon p
   \cr
 \frac1{3\sqrt{2}} \epsilon a(m_1^Re^{-i\phi_1}
 +m_2^Re^{i\phi_1}) &
(1-\frac{1}3 a^2\epsilon)m_2^R &
 i\epsilon q \cr
 -i\epsilon p &i\epsilon q
   &(1-b^2\epsilon)m_3^R \cr}\;,
\ena
where
\bea
p&=&\frac1{\sqrt{6}}b(m_1^Re^{-i\phi_2}-m_3^Re^{i\phi_2})\;,
\nonumber\\
q&=&\frac1{2\sqrt{3}}ab(m_2^Re^{i(\phi_1-\phi_2)}
-m_3^Re^{-i(\phi_1-\phi_2) })\;.
\ena

\vskip 2mm
\noindent
(3-3) The general discussion on the stability

Hereafter, we do not discuss the fully degenerate case, 
$|m_1^R| \simeq |m_2^R|\simeq |m_3^R|$, because this case 
is quite unstable and it is hard to have the definite 
predictions. Therefore, we focus our discussions on 
hierarchical cases; (a) $|m_3^R|>>|m_2^R|>>|m_1^R|$ or 
$|m_3^R|>>|m_1^R|>>|m_2^R|$ and 
(b) $|m_1^R| \simeq |m_2^R| >> |m_3^R|$ or 
$|m_3^R|>>|m_2^R|\simeq |m_1^R|$. 

\vskip 2mm
\noindent
The case (a): The fully hierarchical case

With the use of the analogy of 
the analysis by Haba et al., we expect that all mixing angles 
and the CP violation phase are essentially unchanged 
by the scale change from $M_R$ to $m_Z$. This may be 
simply understood by the consideration that 
the see-saw mechanism can be used to evaluate 
the mixings and the neutrino masses, and thus the 
effect is suppressed by the order of $\epsilon$. 
We checked this result by the numerical computations also. 

\vskip 2mm
\noindent
The case (b): The hierarchical case with $|m_1^R|\simeq |m_2^R|$ 

The situation is slightly complicated in comparison with the 
case (a), because of the degeneracy $|m_1^R|\simeq |m_2^R|$. 
Firstly, we notice that the off diagonal terms are much small 
than $(\tilde m_\nu)_{33}$, or $|(\tilde m_\nu)_{11}|\simeq  
|(\tilde m_\nu)_{22}|$. Therefore, we can use the see-saw 
calculation between $(\nu_1^R,\nu_2^R)$ and $\nu_3^R$, where 
$\nu_i^R$ are mass eigenstates at $M_R$. 
That is, we can safely neglect the contributions from 
$p$ and $q$ terms in the matrix and thus  
we do not need to consider the mixing  between 
$(\nu_1^R,\nu_2^R)$ and $\nu_3^R$. 

Now, the element $V_{i3}$ and $V_{3i}$ 
($i=1,2,3$) at $M_R$ is still valid at 
$m_Z$. That is, $V_{i3}=(V_2)_{i3}$ and 
$V_{3i}=(V_2)_{3i}$. 
As a result, the prediction of 
$\sin^2 2\theta_{atm}>0.999$ 
in Eq.(2)  and the CHOOZ constraint, $|s_{13}|<0.16$ 
are valid at $m_Z$. 

The situation changes depending on the relative sign 
between $m_1^R$ and $m_2^R$. 

\vskip 2mm
\noindent
(b-1) The case where $m_1^R<0$ and $m_2^R>0$

We denote the submatrix for $(\nu_1^R,\nu_2^R)$ 
as $\tilde m_\nu'$ with the approximation $a\simeq 1$ because 
$s'^2<0.04$ is small,  
\bea
\tilde m_\nu' &\simeq&  
\pmatrix{-(1+\Delta-\frac23\epsilon)&
 i\frac{\sqrt{2}}{3} \epsilon \sin \phi_1\cr
i\frac{\sqrt{2}}{3}\epsilon \sin \phi_1 &
1-\frac{1}3 \epsilon\cr}m_2^R\nonumber\\
&=&\pmatrix{1&0\cr0&i\cr}
\pmatrix{-(1+\Delta-\frac23\epsilon)&
\frac{\sqrt{2}}{3} \epsilon \sin \phi_1\cr
\frac{\sqrt{2}}{3}\epsilon \sin \phi_1 &
-(1-\frac{1}3 \epsilon)\cr}m_2^R
\pmatrix{1&0\cr0&i\cr}\;,
\ena
where we defined 
\bea
\Delta=\frac{|m_1^R|-m_2^R}{m_2^R}\;.
\ena

The matrix $\tilde m_\nu'$ is diagonalized by 
\bea
\pmatrix{1&0\cr 0&-i\cr}\pmatrix{c&-s\cr s&c\cr}\;,
\ena
where $c=\cos \theta$ and $s=\sin \theta$ and  
\bea
\tan \theta&=&\frac{\pm\frac{2\sqrt{2}}{3}\epsilon \sin \phi_1}
{\frac 13 \epsilon -\Delta+\sqrt{(\frac 13 \epsilon -\Delta)^2
+\frac89 \epsilon^2 \sin^2 \phi_1}}\;,
\nonumber\\
||m_1|-m_2|&=& m_2^R \sqrt{(\frac 13 \epsilon -\Delta)^2
+\frac89 \epsilon^2 \sin^2 \phi_1} \;,
\ena
and $m_2 \simeq m_2^R$. 

The mixing matrix at $m_Z$ is now obtained by multiplying 
this matrix to $V_2$ in Eq.(11). By looking at the 
structure of $V_2$, we find 
\bea
V_{11}&=&\frac1{\sqrt 3}(c-i\sqrt{2}c's)\;,\nonumber\\
V_{12}&=&\frac1{\sqrt 3}(-s-i\sqrt{2}c'c)\;,
\ena
aside from the irrelevant phases.  
By neglecting the small $s'^2<0.04$, we have $c'= 1$  
and thus we find 
\bea
\sin^2 2\theta_{sol}\simeq \frac89(1+s^2)(1-\frac{s^2}2)\;,
\ena
which takes a value from 8/9 to 1 independent of 
the mixing angle $\theta$. This is due to the phase matrix 
${\rm diag}(1,-i)$ in Eq.(29). 

By the transformation of the matrix in Eq.(29), the 
CP violation phase $\delta$ changes, due to the phase matrix 
${\rm diag}(1,-i)$. The effect is examined by considering the 
Jarlskog parameter which takes the value as
\bea
|J_{CP}|\equiv |{\rm Im}(V_{11}V_{12}^*V_{21}^*V_{22})|
=\frac1{3\sqrt 3} |s'c'(c^2-s^2)|\;,
\ena
and we find
\bea
|\sin \delta | =\frac{|\cos 2\theta |}{
\sqrt{1+\frac18 \sin^2 2\theta \left( 
\frac {\cos 2\theta'}{\cos \theta'}
\right)^2}}\;.
\ena
It should be noted that $\theta=0$  at $M_R$ 
so that $|\sin \delta|=1$. Now we examine the value at 
$m_Z$ from Eq.(34). The angle $\theta$ depends on $\Delta$ which 
is defined in Eq.(28), as we can see in Eq.(30). 
For $\Delta >>\epsilon$ or $\Delta<0$, $|\tan \theta|>>1$ 
or $|\tan \theta|<<1$. Therefore, $|\sin \delta|\sim 1$ 
is realized from Eq.(34). In special cases where 
$\Delta\simeq \epsilon/3$, $\sin \delta$ can become small 
at $m_Z$. In particular, for $\Delta=\epsilon/3$, we find 
$\tan \theta=\pm 1$ and thus we find $\sin \delta=0$. 

Finally, we find
\bea
\Delta_{sol}^2=|m_2^2-m_1^2|\simeq 2m_2^2\sqrt{(\frac 13 \epsilon
-\Delta)^2
+\frac89 \epsilon^2 \sin^2 \phi_1}\;,
\ena
which depends on $m_2$ and $\Delta$. Therefore, we can 
reproduce all three mass squared differences for 
the large angle MSW, the LOW mass and the Just so (Vacuum) 
solutions. For example, when $|\Delta|>>\epsilon$, 
we find $\Delta_{sol}^2\simeq 2m_2^2 \Delta\simeq 
(\Delta_{sol}^2)_{M_R}$, where the value at $M_R$, 
 $(\Delta_{sol}^2)_{M_R}$ is a free parameter 
that we can choose as an input.

\vskip 2mm
\noindent
(b-2) The case where $m_1^R>0$ and $m_2^R>0$

In order to simplify the calculation and to see the 
essence of the analysis, we neglect the term 
$s'^2<0.04$.  Thus we take $a=b=1$ and $\cos \phi_1=1$. 
Then, the submatrix relevant to $\nu_1^R$ and $\nu_2^R$ 
is given by 
\bea
\tilde m_\nu' \simeq  
\pmatrix{(1+\Delta-\frac23\epsilon)&
 \frac{\sqrt{2}}{3} \epsilon \cr
\frac{\sqrt{2}}{3}\epsilon  &
1-\frac{1}3 \epsilon\cr}m_2^R\;.
\ena
After the diagonalization, we find
\bea
m_1&=&\left(1+\frac{\Delta}2-\frac{1}2 \epsilon+{\rm sign}(\Delta)
\frac12\sqrt{D}\right)m_2^R\;,\nonumber\\
m_2&=&\left(1+\frac{\Delta}2-\frac{1}2 \epsilon-{\rm sign}(\Delta)
\frac12\sqrt{ D}\right)
m_2^R\;,
\ena
where ${\rm sign}(\Delta)$ 
takes 1 for $\Delta>0$ and $-1$ for $\Delta<0$ and
\bea
D=(\frac{1}3\epsilon-\Delta)^2
+\frac89 \epsilon^2\;.
\ena
The mass of the third one is 
$m_3=(1-\epsilon )m_3^R$. 
The mixing matrix is 
\bea
\frac1{N} \pmatrix{  {\rm sign}(\Delta)
\sqrt{D}-(\frac 13\epsilon-\Delta) &
  -\frac{2\sqrt{2}}3\epsilon  \cr
  \frac{2\sqrt{2}}3\epsilon  &
 {\rm sign}(\Delta)\sqrt{D}-(\frac13 \epsilon-\Delta) \cr
}\;,
\ena
where $N$ is the normalization factor. 

Now we multiply the above matrix to $V_2$. Aside from the 
unimportant phase and by taking $c'\simeq 1$, we find 
\bea
V_{11}&\simeq&\frac1{\sqrt{3}N}\left\{
{\rm sign}(\Delta)\sqrt{D}-(\frac13 \epsilon-\Delta)
+\frac{4}3\epsilon 
\right\}\;,
\nonumber\\
V_{12}&\simeq&\frac1{\sqrt{3}N}\left\{
-\frac{2\sqrt{2}}3\epsilon +\sqrt{2}\left [
{\rm sign}(\Delta)\sqrt{D}-(\frac13 \epsilon-\Delta)
\right] \right\}\;.
\ena
Now we find
\bea
\sin^2 2\theta_{sol}=\frac89 \left[ 
\frac{\left({\rm sign}(\Delta)\sqrt{D}+\Delta\right)^2 
-\epsilon^2}{\left({\rm sign}(\Delta)\sqrt{D}+\Delta
-\frac{\epsilon}3\right)^2 +\frac 89 \epsilon^2}\right]^2
\;.
\ena

Firstly, since the mass matrix in Eq.(36) is real matrix, 
the CP violation phase $\delta$ are stable and takes 
$\delta=\pi/2$ at $m_Z$. Needless to say, the atmospheric 
neutrino mixing and $s_{13}$ are stable. 

\vskip 2mm
\noindent
(i) The stable $\sin^2 2\theta_{sol}$ 

We focus on the solar neutrino mixing. From Eq.(41), 
we see that if $|\Delta| >>\epsilon$, $\sin^2 2\theta_{sol}
\simeq 8/9$. For $\Delta>0$, this condition is relaxed to 
the condition $\Delta>3\epsilon/2$, where 
$\sin^2 2\theta_{sol}\simeq 8/9$ is realized. 

\vskip 2mm
\noindent
(ii) The unstable $\sin^2 2\theta_{sol}$ 

Now we consider the case where $\sin^2 2\theta_{sol}$ 
becomes small at $m_Z$. We observe that 
$\sin^2 2\theta_{sol}\to 0$ as $\Delta \to 0$. 
This implies that 
$\sin^2 2\theta_{sol}$ becomes small for $\Delta<<\epsilon$, 
while it remains large value for $\Delta>\epsilon$. 

Below, we examine the case $\Delta<<\epsilon$  
to see the $\Delta$ dependence of $\sin^2 2\theta_{sol}$ 
in detail. We expand $\sin^2 2\theta_{sol}$ in terms of 
$\Delta/\epsilon$. We obtain 
\bea
\sin^2 2\theta_{sol}\simeq \frac89 \left(\frac{\Delta}{\epsilon}
\right)^2\;. 
\ena
The small angle which is consistent with the angle for the 
small angle MSW solution, 
$\sin^2 2\theta_{sol}\simeq 10^{-2}$, is realized if we take 
$|\Delta| \sim \frac 1{10}\epsilon$. Next we examine the 
sign of $(m_2^2-m_1^2)\cos 2\theta_{sol}$. 
For $\Delta >0$,  we find  $|V_{11}|>>|V_{12}|$ at $m_Z$, i.e., 
$\cos 2\theta>0$, from Eq.(40).  Then, as we can see from Eq.(37) 
with $m_2^R \simeq m_2$,
\bea
m_1-m_2=m_2\sqrt{(\frac{1}3\epsilon-\Delta)^2
+\frac89 \epsilon^2}>0\;,
\ena
which means 
$m_2^2-m_1^2<0$. Therefore we obtain  
$(m_2^2-m_1^2)\cos 2\theta_{sol}<0$. 
The same conclusion holds for $\Delta <0$ where 
$|V_{11}|<<|V_{12}|$ at $m_Z$ ($\cos 2\theta<0$). That is, 
in both cases, we find $(m_2^2-m_1^2)\cos 2\theta_{sol}<0$. 
It is the standard lore that the small angle MSW solution 
is realized when $(m_2^2-m_1^2)\cos 2\theta_{sol}>0$, 
which is in conflict with our result. 
Recently, Gouv\^ea, Friedland and Murayama[11] have examined 
the case $\cos 2\theta<0$ (dark side) for $m_2^2-m_1^2>0$. 
They found that the region  $\cos 2\theta\sim -0.2$ is 
still possible to explain the solar neutrino problem. 
However, this case corresponds to the large mixing case, 
$\sin^2 2\theta\sim 0.96$ which is not our case. 
In conclusion, when $\Delta\sim \frac{1}{10}\epsilon$, 
$\sin^2 2\theta_{sol}\sim 0.01$ can be realized, but in 
this case the MSW mechanism does not work. 
Therefore, this case is not applicable to solve the solar 
neutrino problem.

\section{Examples -Numerical analysis-}

Since it is hard to search all parameter regions, we set 
$m_2^0=0$ and then varied other parameters, $\tilde m_1$, 
$ m_1^0$ and $ m_3^0$. Here, we exhibit two examples, one
for the stable case where the large angle MSW 
solution for the solar neutrino mixing is realized at $m_{Z}$ 
and Dirac CP phase $\sin \delta$ remain the maximal value,
and the other for the case where  $\sin \delta$ 
becomes to be small at $m_{Z}$. 

\vskip 2mm
\begin{itemize}
\item[(1)] An example for the stable case 

As an example, we adopted input values, 
$(\tilde{m}_1,m_1^0,m_3^0)=(0.0699,-0.0117,-0.025)[{\rm eV}]$ 
which give neutrino masses 
at $M_R$ as 
$(m^R_1,m^R_2,m^R_3)=(0.058200,0.058509,-0.083509)[{\rm eV}]$. 
The values of observables at $M_{R}$ and at $m_{Z}$ are 
given in Table 1, for various values of $\tan \beta$. 

Among various parameters, the parameters relevant to 
atmospheric neutrino mixings, $\Delta_{atm}^2$ and 
$\sin^2 2\theta_{atm}$, $\sin \theta_{13}$ and $\sin \delta$ 
are almost unchanged against the energy scale change for various 
values of $\tan \beta$. The scale dependence for 
$\Delta_{sol}^2$ and $\sin^2 2\theta_{sol}$ depend on the 
values of $\tan \beta$. From the data, 
\bea
0.5<&\sin^2 2\theta_{sol} &<1\;,
\nonumber\\
1\times 10^{-5}[\mbox{eV}^2] <& \Delta_{sol}^2 &<
1\times 10^{-4}[\mbox{eV}^2]\:,
\ena
we have the restriction on $\tan \beta$, 
\bea
\tan \beta =3\sim 13\;, 
\ena
which we can see from Table 1.

\item[(2)] An example to give a small Dirac CP phase at $m_Z$ 

We took input values,  
$(\tilde{m}_1,m_1^0,m_3^0)=(0.3,-0.59651,-0.007)[{\rm eV}]$ 
where neutrino masses 
at $M_R$ are $(m^R_1,m^R_2,m^R_3)=(0.29651,0.29652,-0.30352)
[{\rm eV}]$. We show the values of observables  
at $M_{R}$ and 
at $m_{Z}$ in Table 2, for various values of $\tan \beta$. 

As we can see from Table 2, $\Delta_{atm}^2$, $\sin^2 2\theta_{atm}$ 
and $\sin^2 2\theta_{sol}$ are almost unchanged. 
On the other hand, $\sin \theta_{13}$, $\Delta_{sol}^2$ and 
$\sin \delta$ change depending on $\tan\beta$. In particular, 
$ \sin \delta$ does not change much for small $\tan\beta$, while 
changes substantially for large $\tan\beta$.  
This result is consistent with the discussion given for 
the case $m_1^Rm_2^R<0$ and $\Delta\sim \epsilon/3$.

In Fig.1 and Fig.2, we show the energy scale dependence of 
$m_i^2 (i=1,2)$ and $\sin \delta$ for $\tan\beta=4$ and 10, 
for the parameter set in Table 2. From Fig.1, we see that 
$\Delta_{sol}^2$ increases as the energy scale becomes small and 
also as  $\tan\beta$ becomes large. In Fig.2, we see that 
$\sin \delta$ decreases for both $\tan\beta=4$ and 10. However, 
much faster decrease is observed for the larger $\tan\beta$. 
 
\end{itemize}

\section{Summary and discussions}

We considered the stability of the predictions by  
some special democratic-type neutrino mass matrix, 
which has the quite interesting intrinsic predictions as 
given in Eqs.(1) and (3). We 
assumed that this mass matrix is derived at the right-handed 
mass scale $M_R$ by the see-saw mechanism[2], and then 
considered the mass matrix at the weak scale $m_Z$ and 
its predictions by using the renormalization group. 

We summarize the result as follows:
\begin{itemize}
\item {The case (a)}: The fully hierarchical case
 
This is the case where the neutrino masses at $M_R$ are 
either $|m_3^R|>>|m_1^R|>>|m_2^R|$ 
or $|m_3^R|>>|m_2^R|>>|m_3^R|$. In this case, all predictions 
are stable and  the predictions at $M_R$ are valid at 
$m_Z$.

\item{The case (b)}: The hierarchical case with  $|m_1^R| \simeq
|m_2^R|$ 

If $m_1^R m_2^R<0$, $\sin^2 2\theta_{atm}$ and 
$\sin^2 2\theta_{sol}$ are stable. The CP violation 
phase $\sin \delta$ is also stable for $\Delta>>\epsilon$ or 
$\Delta<0$. For  $\Delta\simeq \epsilon/3$,  $\sin \delta$ 
becomes unstable. 

If $m_1^R m_2^R>0$, $\sin^2 2\theta_{atm}$ and the CP violation 
phase $\delta$ are stable.  The solar mixing angle 
$\sin^2 2\theta_{sol}$ is also stable for $|\Delta|>>\epsilon$. 
For $|\Delta|<\epsilon$, $\sin^2 2\theta_{sol}$ becomes unstable. 
In particular, for $\Delta\simeq \epsilon/10$, 
$\sin^2 2\theta_{sol}$ at $m_Z$ becomes small enough to be 
consistent with the mixing angle for the small angle MSW solution. 
However,  this case 
does not realize the small angle MSW solution. 
\end{itemize}

Our model based on the $Z_3$ symmetry gives quite special 
predictions as given in Eqs.(1) and (3). We emphasize that 
our matrix is intrinsically complex matrix and contains 
the CP violation phase. In particular, our model predicts 
the maximal CP violation phase, which is in contrast to 
most of works where the real neutrino mass matrices are 
treated so that the prediction for the CP violation phase is 
out of reach. The prediction for the CP violation phase in 
the neutrino mass matrix will become a quite important 
topic in view of the near future projects to observe 
the neutrino oscillations, for example, in the neutrino 
factory. 

It is our belief that $Z_3$ symmetry is not only 
useful for describing the neutrino mass matrix, but also 
for the quark mass matrix. The work in this direction will 
be interesting, because we would like to embed the $Z_3$ 
symmetry in the grand unification scheme.

\vskip 5mm
{\Huge Acknowledgment} 
This work is supported in part by 
the Japanese Grant-in-Aid for Scientific Research of
Ministry of Education, Science, Sports and Culture, 
No.12047218.
 
\newpage

\setcounter{section}{0}
\renewcommand{\thesection}{\Alph{section}}
\renewcommand{\theequation}{\thesection .\arabic{equation}}
\newcommand{\apsc}[1]{\stepcounter{section}\noindent
\setcounter{equation}{0}{\Large{\bf{Appendix\,\thesection:\,{#1}}}}}

\apsc{Detailed derivations}

\vskip 2mm
\noindent
(a) The standard parameterization of the mixing matrix 

The particle data group[3] defines the mixing matrix as
\bea
V_{SF}= 
 \pmatrix{c_{12}c_{13}&s_{12}c_{13}&s_{13}e^{-i\delta}\cr
 -s_{12}c_{23}-c_{12}s_{23}s_{13}e^{i\delta}&
  c_{12}c_{23}-s_{12}s_{23}s_{13}e^{i\delta}& s_{23}c_{13}
  \cr
  s_{12}s_{23}-c_{12}c_{23}s_{13}e^{i\delta}&
  -c_{12}s_{23}-s_{12}c_{23}s_{13}e^{i\delta}& c_{23}c_{13}
  \cr}\;.
\ena

\vskip 2mm
\noindent
(b) Diagonalization of $m_\nu(M_R)$ in Eq.(10)

Here, we diagonalize the neutrino mass matrix at $M_R$ and 
thus the predictions are given at $M_R$. 
In order to clarify the property of the democratic-type mass 
matrix, we first transform $m_\nu(M_R)$ by the trimaximal matrix $V_T$ 
\bea
V_T=\frac{1}{\sqrt{3}}\pmatrix{
1 & 1 & 1 \cr
\z & \z^2 & 1 \cr
\z^2 & \z & 1
}\;, 
\ena
where $\z=e^{i2\pi/3}$ ($\z^3=1$) 
and the result is 
\bea
 V_T^T m_\nu (M_R)V_T= \pmatrix{
   m^0_1+\tilde{m}_1&0&0\cr
   0&m^0_2&\tilde m_1\cr
   0&\tilde m_1&m^0_3\cr}\;.
\ena
Then, we transform further by 
\bea
O_1=\pmatrix{1&0&0\cr0&\frac1{\sqrt 2}&-\frac1{\sqrt 2}\cr
           0&\frac1{\sqrt 2} & \frac1{\sqrt 2}\cr}\;,
\ena
and we find 
\bea
(V_TO_1)^T m_\nu (M_R)V_TO_1=
\pmatrix{m_1^0+\tilde m_1&0&0\cr
   0&\tilde m_1+m_2^0+\Delta_- & \Delta_- \cr
  0&\Delta_-&-\tilde m_1 +m_2^0+\Delta_- \cr}
  \;.
\ena 

The matrix $V_1=V_TO_1$ is explicitly given by 
\bea
V_1=\pmatrix{1&0&0\cr0&\z&0\cr 0&0&\z^2}
 \pmatrix{\frac1{\sqrt{3}}&-\sqrt{\frac23}&0\cr 
            \frac{1}{\sqrt 3}&\frac{1}{\sqrt 6}&-\frac{1}{\sqrt 2}\cr
            \frac{1}{\sqrt 3}&\frac{1}{\sqrt 6}&\frac{1}{\sqrt 2}\cr}
   \pmatrix{1&0&0\cr 0&-1&0\cr 0&0&i\cr}\;.
\ena
We have to transform further by 
the orthogonal matrix $O_2$ 
\bea
O_2=\pmatrix{1&0&0\cr 0&c'&-s'\cr 0&s'&c'\cr}\;,
\ena
where $s'$ and $c'$ are defined by Eq.(12). Now the mixing 
matrix is given by $V=V_TO_1O_2$ which is given in Eq.(11).

Below, we give some special cases. 
\vskip 2mm
\noindent
(b-1) The $m_3^0=m_2^0$ case

We have $s'=0$ and $c'=1$ and the mixing matrix is now 
$V=V_1$. Then, the model predicts 
\bea
\sin^2 2\theta_{atm}=1\;,\;\;\sin^2 2\theta_{sol}=\frac89\;.
\ena
There is no CP violation Dirac phase. 

\vskip 2mm
\noindent
(b-2) The $m_2^0=0$ case

The angle $\theta'$ is determined by 
the ratio of $m_2$ and $m_3$, and  we have 
\bea
\sin^22\theta_{sol}=\frac49 \frac{\beta+2}{\beta}\;,
\quad
\sin^2 2\theta_{atm}=\frac49\frac{(\beta+1)(2\beta-1)}{\beta^2}
\;,
\ena
where $\beta=\sqrt{|m_2/m_3|}+\sqrt{|m_3/m_2|}\ge 2$. If 
$\beta$ is close to 2, we have the large solar neutrino mixing 
and also the large atmospheric neutrino mixing.

\begin{center}
\begin{table}
\begin{tabular}{|c|c|c|c|c|c|c|}\hline
\multicolumn{7}{|c|}{Values at $M_R$ scale}\\ \hline
&$\sin\theta_{13}$&$\Delta^2_{atm}[\mbox{eV}^2]$&$\sin^22\theta_{atm}$
&$\Delta^2_{sol}[\mbox{eV}^2]$&$\sin^22\theta_{sol}$&$\sin\delta$\\
\hline
&0.072148&$3.5865\times 10^{-3}$&0.99997&$3.6048\times
10^{-5}$&0.88195&1\\ \hline
\multicolumn{7}{|c|}{Values at $m_Z$ scale}\\ \hline
$\tan\beta$&$\sin\theta_{13}$&$\Delta^2_{atm}[\mbox{eV}^2]$&$\sin^22\theta_{atm}$&
$\Delta^2_{sol}[\mbox{eV}^2]$&$\sin^22\theta_{sol}$&$\sin\delta$\\
\hline
3&0.072149&$3.5837\times 10^{-3}$&0.99998&$3.6422\times 
10^{-5}$&0.86307&1.0000 \\ \hline
4&0.072150&$3.5818\times 10^{-3}$&0.99999&$3.6702
\times 10^{-5}$&0.84934&1.0000 \\ \hline
5&0.072151&$3.5793\times 10^{-3}$&0.99999&$3.7084
\times 10^{-5}$&0.83116&1.0000 \\ \hline
6&0.072152&$3.5763\times 10^{-3}$&1.0000&$3.7584
\times 10^{-5}$&0.80832&1.0000 \\ \hline
7&0.072153&$3.5728\times 10^{-3}$&1.0000&$3.8218
\times 10^{-5}$&0.78068&1.0000 \\ \hline
8&0.072155&$3.5688\times 10^{-3}$&0.99999&$3.9006
\times 10^{-5}$&0.74831&1.0000 \\ \hline
9&0.072156&$3.5643\times 10^{-3}$&0.99999&$3.9968
\times 10^{-5}$&0.71149&1.0000 \\ \hline
10&0.072158&$3.5594\times 10^{-3}$&0.99998&$4.1123
\times 10^{-5}$&0.67078&1.0000 \\ \hline
11&0.072160&$3.5540\times 10^{-3}$&0.99996&$4.2491
\times 10^{-5}$&0.62696&1.0000 \\ \hline
12&0.072162&$3.5481\times 10^{-3}$&0.99993&$4.4087
\times 10^{-5}$&0.58102&1.0000 \\ \hline
13&0.072164&$3.5418\times 10^{-3}$&0.99988&$4.5926
\times 10^{-5}$&0.53404&1.0000 \\ \hline
14&0.072166&$3.5351\times 10^{-3}$&0.99982&$4.8021
\times 10^{-5}$&0.48711&1.0000 \\ \hline
15&0.072168&$3.5280\times 10^{-3}$&0.99974&$5.0380
\times 10^{-5}$&0.44124&1.0000 \\ \hline
\end{tabular}
\caption{The predicted values of various observable at $m_Z$.
As input values at $M_R$, we choose
$(\tilde{m}_1,m_1^0,m_3^0)=(0.0699, -0.0117, -0.025)$ which are
equivalent to the choice of neutrino masses at $M_R$
$(m^R_1,m^R_2,m^R_3)=(
0.058200,0.058509,-0.083509)\mbox[eV]$.In this case , observable are
almost stable.
However  $\sin^2 2\theta_{sol}$ becomes smaller as $\tan\beta $
becomes larger.  
} 
\end{table}
\end{center}

\newpage
\begin{center}
\begin{table}
\begin{tabular}{|c|c|c|c|c|c|c|}\hline
\multicolumn{7}{|c|}{Values at $M_R$}\\ \hline
&$\sin\theta_{13}$&$\Delta^2_{atm}[\mbox{eV}^2]$&$\sin^22\theta_{atm}$
&$\Delta^2_{sol}[\mbox{eV}^2]$&$\sin^22\theta_{sol}$&$\sin\delta$\\
\hline
&0.004763&$4.2065\times 10^{-3}$&1.0000&$6.1770\times 10^{-6}$
&0.88886&1 \\ \hline\hline
\multicolumn{7}{|c|}{Values at $m_Z$}\\ \hline
$\tan\beta$&$\sin\theta_{13}$&$\Delta^2_{atm}[\mbox{eV}^2]$&$\sin^22\theta_{atm}$&
$\Delta^2_{sol}[\mbox{eV}^2]$&$\sin^22\theta_{sol}$&$\sin\delta$\\
\hline
3&0.005870&$4.1939\times 10^{-3}$&0.99997&$1.5964\times
10^{-5}$&0.88887&0.82400 \\ \hline
4&0.007538&$4.1851\times 10^{-3}$&0.99993&$2.2724\times
10^{-5}$&0.88886&0.65154 \\ \hline
5&0.001015&$4.1740\times 10^{-3}$&0.99984&$3.1305\times
10^{-5}$&0.88881&0.49382 \\ \hline
6&0.013674&$4.1606\times 10^{-3}$&0.99968&$4.1619\times
10^{-5}$&0.88871&0.37619 \\ \hline
7&0.018061&$4.1451\times 10^{-3}$&0.99941&$5.4560\times
10^{-5}$&0.88853&0.29373 \\ \hline
8&0.023293&$4.1275\times 10^{-3}$&0.99900&$6.6999\times
10^{-5}$&0.88825&0.23605 \\ \hline
9&0.029372&$4.1080\times 10^{-3}$&0.99839&$8.1782\times
10^{-5}$&0.88783&0.19496 \\ \hline
10&0.036315&$4.0869\times 10^{-3}$&0.99755&$9.7732\times
10^{-5}$&0.88722&0.16505 \\ \hline
11&0.044144&$4.0643\times 10^{-3}$&0.99640&$1.1464\times
10^{-4}$&0.88639&0.14282 \\ \hline
\end{tabular}
\caption{The predicted values of various obsarvable at $m_Z$.
As input values at $M_R$, we choose 
$(\tilde{m}_1,m_1^0,m_3^0)=(0.3, -0.59651,-0.007)$ 
which are equivarent to the choice of neutrino masses 
at $M_R$ $(m^R_1,m^R_2,m^R_3)=(-0.29651,0.29652,-0.30352)\mbox[eV]$.
Note that relative sign of $m^R_1$ and $m^R_2$ is mainus.
In this case, $\sin\delta $ becomes to be small.
}
\end{table}
\end{center}

\begin{figure}
\begin{center}
\includegraphics[scale=0.7]{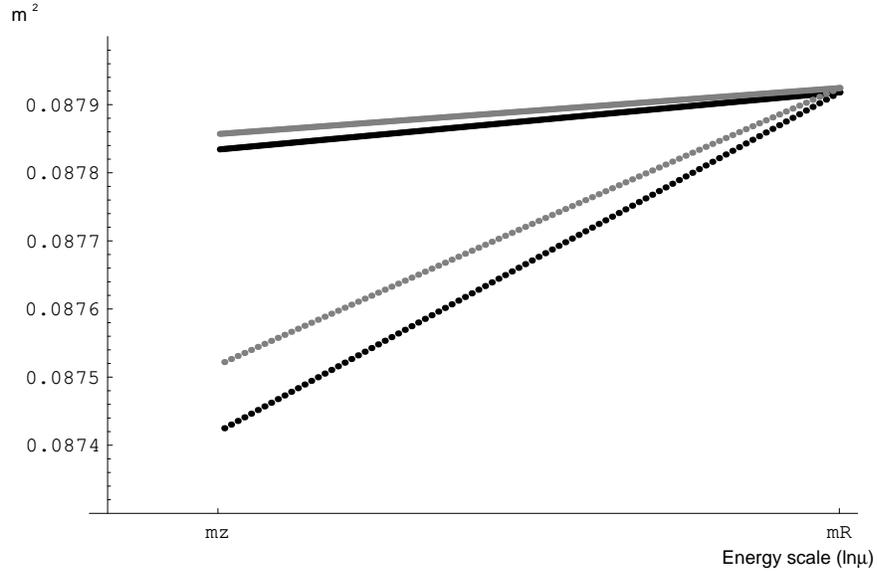}
\end{center}
\caption{Energy scale dependence of 1st and 2nd generation's neutrino 
masses $[\mbox{eV}^2]$(squared masses) for the parameter set 
at $M_R$ given in Table 2($m^R_1\cdot m^R_2 < 0$).  
Black line (dots) is for $m_1^2$ and gray line
(dots) is for $m_2^2$. Solid line is for $\tan \beta =4$ and
dashed line is for $\tan\beta =10$. The horizontal axes describe 
energy scale($\log \mu$).
}
\end{figure}

\begin{figure}
\begin{center}
\includegraphics[scale=0.7]{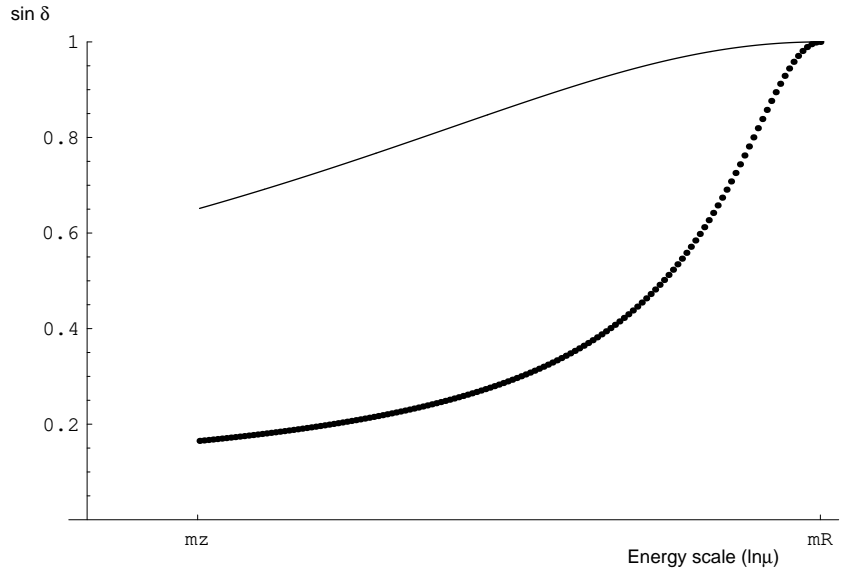}
\end{center}
\caption{Energy dependence of $\sin\delta$ for the 
parameter set at $M_R$ given in Table 2 
($m^R_1\cdot m^R_2 < 0$) and this case shows small CP violation
angle at $m_Z$, while it is large at $M_R$.
Solid curves correspond to $\tan\beta =4$ while dashed are 
for $\tan\beta =10$.}
\end{figure}

\end{document}